\documentclass[sn-mathphys-num]{sn-jnl}
\usepackage{graphicx}%
\usepackage{multirow}%
\usepackage{amsmath,amssymb,amsfonts}%
\usepackage{amsthm}%
\usepackage{mathrsfs}%
\usepackage[title]{appendix}%
\usepackage{xcolor}%
\usepackage{textcomp}%
\usepackage{manyfoot}%
\usepackage{booktabs}%
\usepackage{algorithm}%
\usepackage{algorithmicx}%
\usepackage{algpseudocode}%
\usepackage{listings}%
\usepackage{graphicx}% Include figure files
\usepackage{dcolumn}% Align table columns on decimal point
\usepackage{bm}% bold math
\usepackage[export]{adjustbox}
\usepackage{xcolor}
\usepackage{MnSymbol}
\usepackage{footmisc}
\theoremstyle{thmstyleone}%
\theoremstyle{thmstyletwo}%
\theoremstyle{thmstylethree}%

\newcommand{\bra}[1]{\left\langle{#1}\right|}
\newcommand{\ket}[1]{\left|{#1}\right\rangle}

\begin{document}

\title{Bose-Hubbard model in the presence of artificial magnetic fields: Ground state and thermal phase diagrams}

\author{\fnm{Mohammadamin} \sur{Jaberi}}\email{}

\author{\fnm{Fatemeh} \sur{Heydarinasab}}\email{fheydari@phys.usb.ac.ir}

\affil{\orgdiv{Department of Physics}, \orgname{Faculty of Science, University of Sistan and Baluchestan}, \city{Zahedan}, \country{Iran}}

\date{\today}
 
%%==================================%% abstract %%
%%==================================

\abstract{We consider effects of artificial magnetic fields on the ground state of the two-dimensional Bose-Hubbard model. Using an asymmetric Bose-Hubbard model, we demonstrate that the frustrating hopping energy localizes bosons and enlarges insulators and supersolid borders. We show these increments in the presence of artificial gauge fields up to a symmetric point of the field. 
Moreover, our calculations exhibit the real space modulations of the superfluid and supersolid phases. 
The bosonic current exhibits vortices in these phases, which their configurations depend on the commensuration between the magnetic field and the lattice. 
Although the magnetic field breaks the translational symmetry of the square lattice explicitly, this symmetry is restored in the Mott insulator phase. The filling factor exhibits a checkerboard pattern in the density-wave and supersolid phases, regardless of the magnetic field strength. 
We explore thermal fluctuations and demonstrate the robustness of insulators and SS phases up to temperatures comparable to the interaction energy, which support the feasibility of observing such phases in experiments.
}


\maketitle

\section{introduction}\label{intro}

Optical lattices are clean and highly controllable systems that provide experimental realization of quantum phases \cite{greiner2002quantum, bloch2008many, lewenstein2012ultracold}. 
The versatility of adjustable parameters makes them a suitable playground for simulating quantum phases that are hardly accessible in solid-state systems. 

Bose-Hubbard model (BHM) is the widely used model to simulate quantum phase transitions in the optical lattices \cite{greiner2002quantum, stanescu2010topological}. 
Recent experiments promise the emergence of topological phases by this model. 
The Coriolis force generated by a rotating reference frame mimics the Lorentz force in the charged particles systems \cite{cooper2008rapidly, fetter2009rotating, schweikhard2007vortex, williams2010observation}. 
As this magnetic field is generated in the charge neutral systems, it is called artificial magnetic field (AMF).

Laser-induced tunneling is another way for the realization of the artificial magnetic fields in the bosonic systems \cite{spielman2009raman, lin2009synthetic, atala2014observation, aidelsburger2011experimental, aidelsburger2013realization}. 
These synthetic fields induce Peierls phase factors in the hopping terms of the BHM, and leads to modifications of the single-particle spectrum and the emergence of nontrivial spatial structures \cite{powell2011bogoliubov}. 

The effective magnetic field which is generated in these systems is large enough to reach the quantum Hall regime \cite{furukawa2013integer, sorensen2005fractional, hafezi2007fractional, moller2009composite, hormozi2012fractional, palmer2008optical}. 
The ability to control dimensionality and interaction parameters, in addition to tunability of the magnetic field, make these systems more interesting for generating integer \cite{wilkin2000condensation, cooper2001quantum, rezayi2005incompressible} and half integer \cite{sorensen2005fractional, hafezi2007fractional, hormozi2012fractional} quantum Hall effects. 
Moreover, due to the confinement of atoms in a small volume, the interaction energies and the excitation gap are larger in trapped atoms, making the realization of the fractional quantum Hall effect is easier \cite{hafezi2007fractional}.

Theoretically, artificial magnetic field enlarges the Mott-insulator phase in the Bose-Hubbard model \cite{oktel2007mean, sachdeva2010cold, iskin2012artificial, kelecs2015mott, sachdeva2017extended}. This enlargement is related to the localization of the single-particle dynamics even for non-interacting systems \cite{iskin2012artificial}. Moreover, AMF leads to superfluid phase with simultaneous spatial order which its configuration depends on commensuration between the magnetic field and the lattice \cite{powell2011bogoliubov}. 

Nearest-neighbor (NN) interaction, which is the result of dipole-dipole interaction \cite{baier2016extended}, would break translational symmetry in the extended Bose-Hubbard model (eBHM) \cite{heydarinasab2017inhomogeneous, heydarinasab2018spin, yamamoto2012quantum}. Therefore, density-wave (DW) and supersolid (SS) phases are added to the phase diagram of this model  \cite{heydarinasab2017inhomogeneous, heydarinasab2018spin, yamamoto2012quantum}. SS phase is the result of coexistence of translational and rotational symmetry breakings \cite{leggett1970can, kim2004probable}. Realization of this phase was on debates \cite{boninsegni2006fate} for many years. Including AMF to the eBHM increases SS regions which ease observing this phase in experiments. Moreover, excess particles in the eBHM with NN repulsion represent bosonic analog of the fractional quantum Hall effect at a particular filling factors  \cite{kuno2017bosonic}.

In this manuscript, we provide new insights into the effects of AMF on the ground state phase diagrams of the two-dimensional BHM and eBHM. At first, single-particle spectrum of the BHM+AMF is considered that shows the Hofstadter-butterfly structure and becomes degenerate corresponding to the size of the field. In order to predict effects of AMF in the presence of interactions, we use asymmetric BHM and eBHM. These results prove that the frustrating hopping energies enlarge the insulator and SS borders. Effects of different ranges of AMF on the BHM and eBHM is considered and showed the increment of the insulators and SS borders with AMF up to symmetric point of the field. 
We consider evolution of the symmetries and order parameters in the response of different ranges of interactions and magnetic field thoroughly. 
Although AMF breaks translational symmetry of the square lattice explicitly, this symmetry returns to the system in the MI phase. We show checkerboard pattern of the filling factor in the DW and SS phases. However, there are real space modulation of the bosonic current in the SF and SS phases corresponding to the size of the magnetic field. 
Finally, we demonstrate stability of different phases up to temperatures which is comparable with interaction terms. 
Up to our knowledge the effects of AMF on the properties of
SS phase was not considered elsewhere. 

This paper is organized as follows. In the section \ref{sec:model}, we introduce the model and include AMF as a phase factor in the hopping energy. In the section \ref{sec:Ek}, we find single-particle spectrum in different ranges of magnetic fields, and illustrate that this phase factor breaks translational symmetry of the square lattice which modulate the superfluid order parameter. Method is introduced in the section \ref{sec:method}.
In order to anticipate effects of frustrated hopping energy in the presence of AMF, phase diagrams of the BHM and eBHM with asymmetric hopping energies are plotted in the section \ref{sec:ABH}, and increment of the insulators and supersolid phases are shown. 
In the section \ref{sec:BHM+AMF}, using different order parameters, we find ground state phase diagrams in different ranges of interactions. 
Effects of thermal fluctuations is considered in the section \ref{sec:temp} and concluding remarks are stated in the section \ref{sec:conclusions}.

%%%%%%%%%%%%%%%%%%%%%%%%%%%%%%%%%%%%%%%%%
%%%%%%%%%%%%%%%%% model %%%%%%%%%%%%%%%%%
%%%%%%%%%%%%%%%%%%%%%%%%%%%%%%%%%%%%%%%%%

\section{model}\label{sec:model}

We consider Bose-Hubbard model on the two-dimensional square lattice in the presence of artificial magnetic field. 
Previous experiments simulated effective Landau gauge potentials, in which the vector potential was parallel to one of the square lattice axes \cite{lin2009bose, lin2009synthetic}. We take the vector potential along $\hat{y}$ direction optionally:
%%%%%%%%%%%%%%%%%%%%%%%%%%%%%%
\begin{eqnarray}\label{eq:model}
  \nonumber  H = \sum_{i} &  [~ -& t ~ (b^\dag_{i+\hat{x}} b_{i} + e^{- \mathrm{i} 2 \pi \alpha x_i} ~ b^\dag_{i+\hat{y}} b_{i} + h.c.) \\
    &+& \frac{U}{2} ~ n_{i} (n_{i} -1) ~ - ~ \mu ~ n_{i}],
\end{eqnarray}
%%%%%%%%%%%%%%%%%%%%%%%%%%%%%%
where $b^\dag_{i} (b_{i})$ is the creation (annihilation) operator of a boson at the lattice site $i$, and $n_{i} = b^\dag_{i} b_{i}$ is the boson number operator. 
$U$ is on-site interaction energy which is set as the scale of energy ($U=1$). $t$ and $\mu$ are hopping amplitude between nearest neighbors and chemical potential, respectively. Moreover, $2 \pi \alpha$ is the dimensionless flux per plaquette which defines strength of the magnetic field with $\alpha = \rcircleleftint \mathrm d \mathbf{r} \cdot \mathbf{A}(\mathbf{r})$ where $\mathbf{A}$ is the vector potential that is related to the synthetic magnetic field as  $\mathbf{B} = \nabla \times \mathbf{A}$. 
Here $\alpha=\frac{1}{q}$, which $q$ is integer and specifies period of the AMF on the lattice. This proves that magnetic field breaks translational symmetry along $\hat{x}$ direction explicitly, and the unit cell becomes $q \times 1$. Therefore, in the condensed phases, we will see this specific modulation on the square lattice.  
On the other hand, as the number operator $n_i$ is gauge invariant, translational symmetry would be reserved in the insulator phases with large on-site interaction $U$. Also, the system displays checkerboard pattern in the DW phase with large NN interactions. 

%%%%%%%%%%%%%%%%%%%%%%%%%%%%%%%%%%%%%%%%%
%%%%%%%%%%%%%%%% spectrum %%%%%%%%%%%%%%%
%%%%%%%%%%%%%%%%%%%%%%%%%%%%%%%%%%%%%%%%%

\section{Single-particle spectrum}\label{sec:Ek}

Including AMF in the Bose-Hubbard model (Eq. \ref{eq:model}) explicitly breaks translational symmetry in the $\hat{x}$ direction and enlarges the unit cell to $q \times 1$ for $\alpha = 1/q$. Therefore, the magnetic field breaks dispersion relation to $q$ subbands, and the first Brillouin zone reduces to $k_x \in \{\frac{-\pi}{q} : \frac{\pi}{q} \}$ and $k_y \in \{-\pi : \pi \}$. 

In order to find effect of AMF on the excitation spectrum of the BHM, in this section we calculate single-particle ($U=0$) spectrum of this model. 
Using Fourier transformation of the creation and annihilation operators at site $i$:
%%%%%%%%%%%%%%%%%%%%%%%%%%%%%%%%%
\begin{eqnarray}
    \nonumber b_{i} &=& \frac{1}{N} \sum_{\boldsymbol{k}} e^{\mathrm{i} \boldsymbol{k} r_i} a_{\boldsymbol{k}}  \\
    b^\dag_{i} &=& \frac{1}{N} \sum_{\boldsymbol{k}} e^{-\mathrm{i} \boldsymbol{k} r_i} a^\dagger_{\boldsymbol{k}}, 
\end{eqnarray}
%%%%%%%%%%%%%%%%%%%%%%%%%%%%%%%%
we transform the Hamiltonian in Eq. \ref{eq:model} into momentum space:
%%%%%%%%%%%%%%%%%%%%%%%%%%%%%%%%%%  
\begin{eqnarray}
H(\boldsymbol{k}) = -t ~\begin{pmatrix}
    \varepsilon_0 &  e^{\mathrm{i} k_x} & 0 & \cdots & e^{-\mathrm{i} k_x} \\
    \\
    e^{-\mathrm{i} k_x} & \varepsilon_1 & e^{\mathrm{i} k_x} & \ddots & 0 &  \\
    \\
    0 & e^{-\mathrm{i} k_x} & \varepsilon_2 & \cdots & 0 &  \\
    \\
    \vdots & \vdots & \ddots &  \ddots & \vdots \\
    \\
    e^{\mathrm{i} k_x} & 0 &  \cdots & e^{-\mathrm{i} k_x} & \varepsilon_{q-1},
\end{pmatrix},
\end{eqnarray}
%%%%%%%%%%%%%%%%%%%%%%%%%%%%%%%%%%
where $N$ is the total number of bosons and $\varepsilon_q = 2 cos(k_y + 2 \pi \alpha (q-1)) + \mu/t$. Para-diagonalization \cite{colpa1978diagonalization} of this relation results excitation spectrum of the BHM+AMF.
%%%%%%%%%%%%%%%%%%%%%%%%%%%%%%%%%%
\begin{figure}[t]
\centering
\includegraphics[scale=0.5]{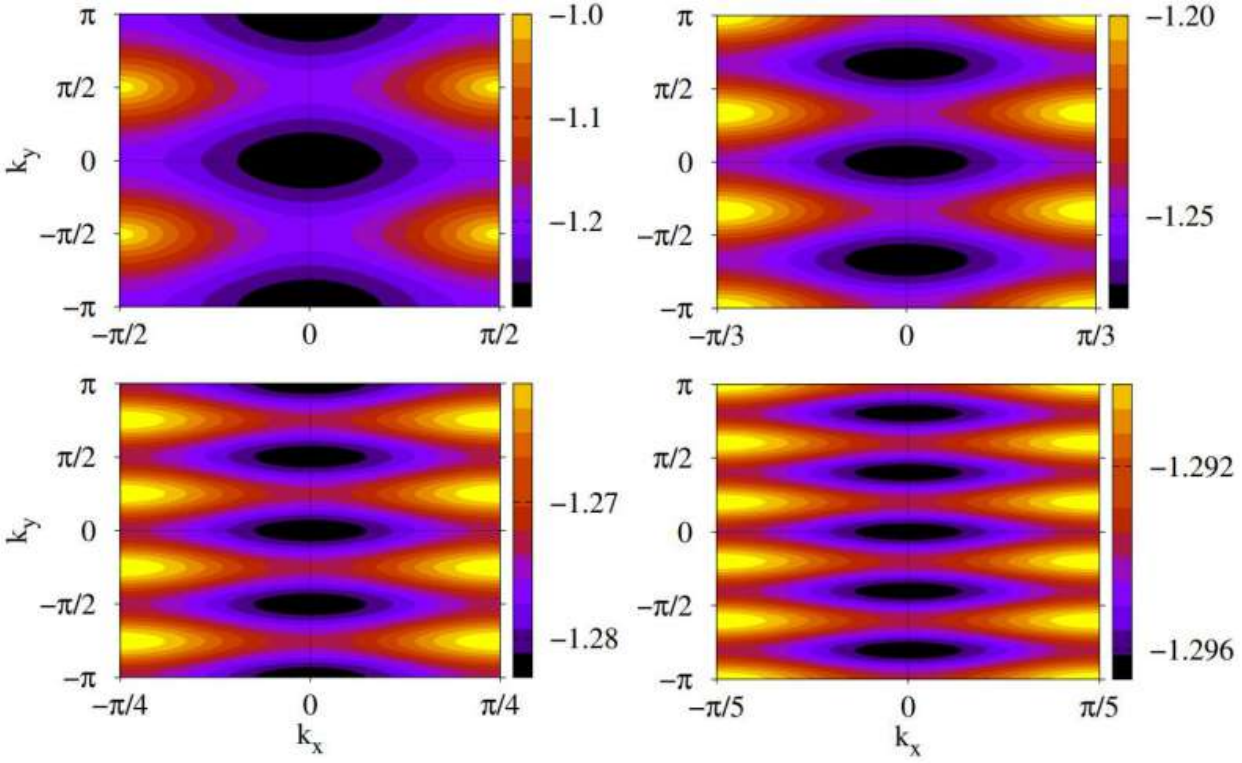}
\caption{Lowest band of the non-interacting BHM+AMF with different gauge fields: $\alpha = \frac{1}{2}$ (top left), $\frac{1}{3}$ (top right), $\frac{1}{4}$ (bottom left), and $\frac{1}{5}$ (bottom right).}
\label{fig:Ek}
\end{figure}
%%%%%%%%%%%%%%%%%%%%%%%%%%%%%%%%%%%%%%%%%

In the Fig. \ref{fig:Ek}, we have plotted lowest single-particle spectrum of the BHM+AMF in the SF phase. This figure proves that the lowest band structure is degenerate for $k_y$s which are separated by $2 \pi /q$ values. Therefore, Brillouin zone would be decreased to $\{\frac{-\pi}{q} : \frac{\pi}{q} \}$ for the both directions. 
It means that in the SF phase order parameters represent this specific modulation along $\hat{y}$ direction too. Using order parameters, we prove this fact in the next sections.

%%%%%%%%%%%%%%%%%%%%%%%%%%%%%%%%
\begin{figure}
\centering
\includegraphics[scale=0.38]{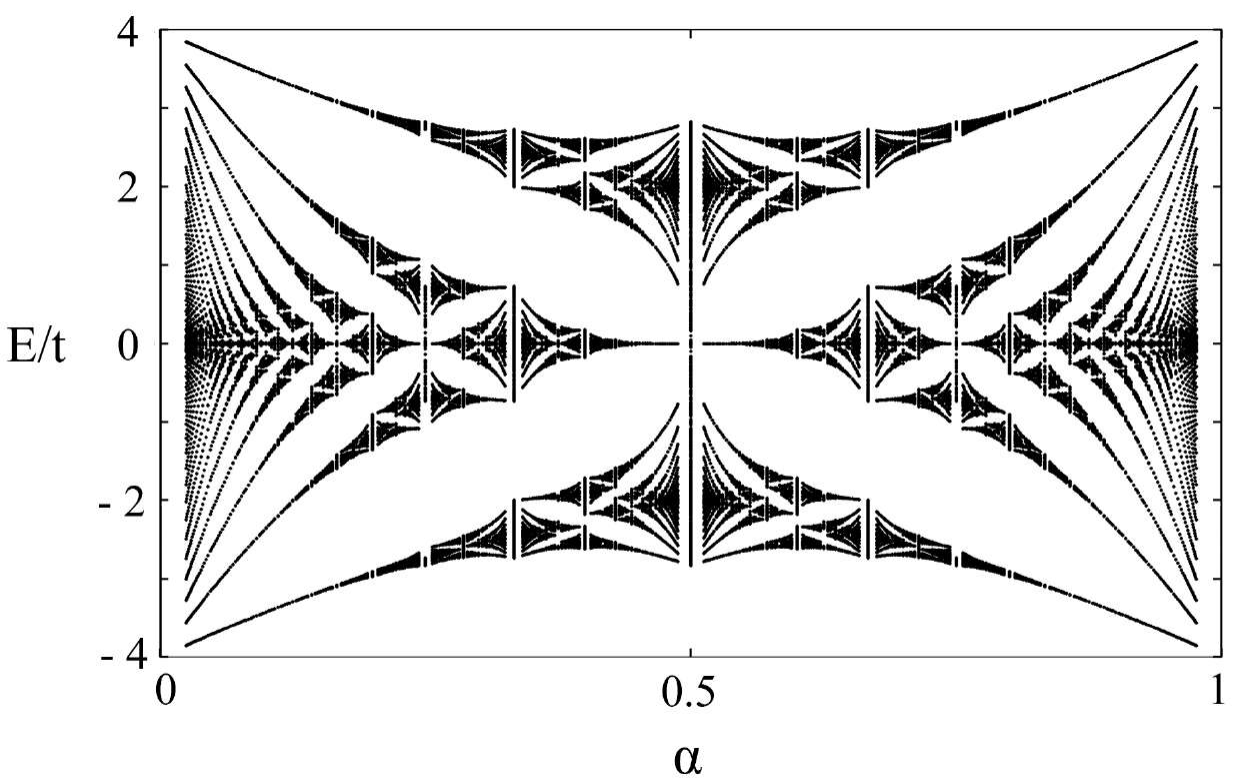}
\caption{The Hofstadter-butterfly \cite{hofstadter1976energy}: The single-particle energies E as a function of flux $\alpha$.}
\label{fig:butterfly}
\end{figure}

The single-particle energy of this model exhibits a fractal structure as a function of $\alpha$, known as Hofstadter-butterfly \cite{hofstadter1976energy} and is symmetric around $\alpha = 1/2$ for $0 \leq \alpha \leq 1$, see Fig. \ref{fig:butterfly}. Therefore, we expect that the results on two sides of the symmetric point are the same. For example, we verified (not shown) that the phase diagrams of the $\alpha=1/5$ is as same as $4/5$, other than a reversal in the direction of the bosonic current evidently. So, we have limited our discussion to magnetic fields up to $1/2$ for simplicity.

%%%%%%%%%%%%%%%%%%%%%%%%%%%%%%%%%%%%%%%%%

%%%%%%%%%%%%%%%%%%%%%%%%%%%%%%%%%%%%%%%%%
%%%%%%%%%%%%%%% GW method %%%%%%%%%%%%%%%
%%%%%%%%%%%%%%%%%%%%%%%%%%%%%%%%%%%%%%%%%

\section{method}\label{sec:method}

We use the single-site Gutzwiller mean-field (SGMF) method \cite{rokhsar1991gutzwiller, krauth1992gutzwiller, colussi2022quantum, luhmann2013cluster} to find ground state phase diagrams. In this method the creation and annihilation operators decompose as:
%%%%%%%%%%%%%%%%%%%%%%%%%%%%%%
\begin{eqnarray}    
\nonumber b_{i} &=& \phi_{i}+\delta b_{i} \\
b^\dag_{i} &=& \phi_{i}^\ast+\delta b^\dag_{i},
\end{eqnarray}
%%%%%%%%%%%%%%%%%%%%%%%%%%%%%%
at a particular lattice site $i$ respectively. The mean-field part $\phi_{i}=\langle b_{i} \rangle$ is SF order parameter and $\phi_{i}^\ast$ is its complex conjugate that are defined with respect to the ground state of the system. $\delta b_{i}$ ($\delta b^\dag_{i}$) is quantum fluctuations around mean-field results. 
 
Using these relations the Hamiltonian in Eq. (\ref{eq:model}) is reduced to the mean-field Hamiltonian:
%%%%%%%%%%%%%%%%%%%%%%%%%%%%%%%%%
\begin{eqnarray}
H^{MF} &=& \sum_{i} H^{MF}_{i},
\end{eqnarray}
%%%%%%%%%%%%%%%%%%%%%%%%%%%%%%%%%%%%%%
up to linear quantum fluctuations. $H^{MF}_{i}$ is the single-site mean-field Hamiltonian: %%%%%%%%%%%%%%%%%%%%%%%%%%%%%%
\begin{eqnarray} \label{eq:MFH}
\nonumber && H^{MF}_{i} = \\
\nonumber && -t ~ b_{i} ~( \phi^\ast_{i+\hat{x}} + \phi^\ast_{i-\hat{x}} + e^{-\mathrm{i} 2 \pi \alpha x_i} \phi^\star_{i+\hat{y}} + e^{\mathrm{i} 2 \pi \alpha x_i} \phi^\star_{i-\hat{y}})+h.c. \\
 && + ~ \frac{U}{2} ~ n_{i} (n_{i} -1) -\mu ~ n_{i}.
\end{eqnarray}
%%%%%%%%%%%%%%%%%%%%%%%%%%%%%%

The ground state of the system, $\ket \Psi_{G}$, is defined as a direct product of the normalized single-site ground states $\ket \psi_{i}$ which is written in the Fock space:
%%%%%%%%%%%%%%%%%%%%%%%%%%%%%%%%%%
\begin{eqnarray}
\ket \Psi_{G} = \prod_{i} \ket \psi_{i} = \prod_{i} ~ \sum_{n=0}^{n_b} c_{n}^{i} \ket n_{i}.
\end{eqnarray}
%%%%%%%%%%%%%%%%%%%%%%%%%%%%%%%%
$n_b$ is the maximum allowed occupation number, and $c_{n}^{i}$ are the coefficients of the occupation state $\ket n_{i}$ at lattice sites $i$. The ground state is computed from the self-consistent solution of the single-site energy. 

Using these relations, the superfluidity is defined as:
%%%%%%%%%%%%%%%%%%%%%%%%%%%%%%%%%%%%%%%
\begin{eqnarray}
\phi_{i}= \bra{\Psi_{G}} b_{i} \ket{\Psi_{G}} = 
\sum_{n=0}^{n_b} ~ c_{n-1}^ {i \ast} ~ c_{n}^ {i} ~ \sqrt{n}.
\end{eqnarray} 
%%%%%%%%%%%%%%%%%%%%%%%%%%%%%%%%%%%%%%%
According to this expression, it is clear that $\phi_{i}$ is zero in the DW phase where only one occupation number state is occupied on every lattice sites and one non-zero coefficient $c^{i}_n$ contributes to the single-site wave function. Moreover, $\phi_{i}$ vanishes in the MI phase, where these coefficients are identical across all lattice sites. But $\phi_{i}$ is finite in the SF and SS phases with number fluctuations. 
The occupancy and number fluctuation at every lattice sites are defined as:
%%%%%%%%%%%%%%%%%%%%%%%%%%%%%%%%%%%%%%%
\begin{eqnarray}\label{eq:density}
\langle n_{i} \rangle = \sum_{n=0}^{n_b} ~ |{c}^{i}_{n}|^2 ~ n,
\end{eqnarray} 
%%%%%%%%%%%%%%%%%%%%%%%%%%%%%%%%%%%%%%%%
and
%%%%%%%%%%%%%%%%%%%%%%%%%%%%%%%%%%%%%%%%
\begin{eqnarray}
\delta n _{i} = \sqrt{\langle n^2_{i} \rangle -\langle n_{i} \rangle^2 },
\end{eqnarray} 
%%%%%%%%%%%%%%%%%%%%%%%%%%%%%%%%%%%%
respectively. The plateaus of the order parameter $\langle n_{i} \rangle$ can differentiate insulator phases from SF and SS phases.
Local compressibility $\delta n_{i}$ is zero in the insulator phases which makes them incoherent. While it's large in the SF phase which make it a strong phase coherent phase. Due to SF component, this order parameter is finite in the SS phase too. 

%%%%%%%%%%%%%%%%%%%%%%%%%%%%%%%%%%%%%%%%%
%%%%%%%%%%%%%%%% spectrum %%%%%%%%%%%%%%%
%%%%%%%%%%%%%%%%%%%%%%%%%%%%%%%%%%%%%%%%%

\section{Asymmetric Bose-Hubbard model} \label{sec:ABH}

The inclusion of a magnetic field in the Bose-Hubbard model frustrates hopping energies along different directions. In order to find initial insight into this effect, we consider the standard Bose-Hubbard model with asymmetric hopping amplitudes, such that hopping energy along the $\hat{y}$ direction is smaller than the $\hat{x}$ direction:
%%%%%%%%%%%%%%%%%%%%%%%%%%%%%%
\begin{eqnarray}\label{eq:ABH}
  \nonumber  H = \sum_{i} & [ ~ -& t ~ (b^\dag_{i+\hat{x}} b_{i} + \frac{1}{2} ~ b^\dag_{i+\hat{y}} b_{i} + h.c.) \\
    &+& \frac{U}{2} ~ n_{i} (n_{i} -1) ~ - \mu  ~ n_{i} ],
\end{eqnarray}
%%%%%%%%%%%%%%%%%%%%%%%%%%%%%%

%%%%%%%%%%%%%%%%%%%%%%%%%%%%%%
\begin{figure}[t]
\centering
\includegraphics[scale=0.6]{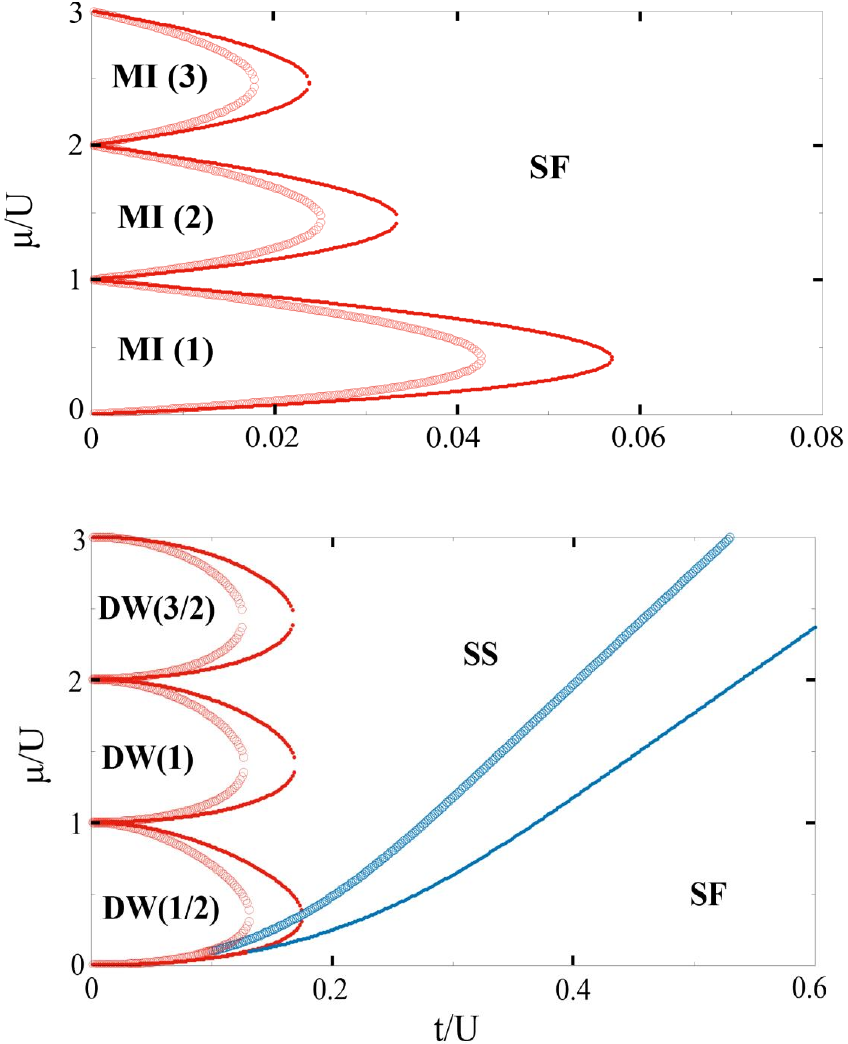}
\caption{SGMF $t-\mu$ ground state phase diagrams in the units of on-site interaction ($U=1$) for top: BHM and bottom: eBHM with NN interaction $V_1/U=0.5$, and symmetric (circle) and asymmetric (dot) hopping energies. MI-SF and DW-SS borders are plotted with red color and SS-SF borders with blue.}
\label{fig:PD-ABH}
\end{figure}
%%%%%%%%%%%%%%%%%%%%%%%%%%%%%%%%%%%%%%%%

Ground state phase diagram of this model together with the standard BHM is plotted in the top panel of the Fig. \ref{fig:PD-ABH}. 

In the absence of hopping energy $t$, the incompressible Mott insulator appears, in which translational and rotational symmetries are preserved. The average boson number exhibits a plateau in this phase, while filling factor is increased by $\mu$ in different MI lobes. 

As the hopping energy $t$ increases, the system undergoes a phase transition where the U(1) symmetry is spontaneously broken and a superfluid phase emerges.
Asymmetric hopping energy violates coherency of bosonic motions which enlarges the insulator borders in comparison with the standard symmetric BHM. Therefore, MI phases are broadened in the asymmetric BHM.

In the bottom panel of Fig. \ref{fig:PD-ABH}, nearest-neighbor interaction:
%%%%%%%%%%%%%%%%%%%%%%%%%%%%%%
\begin{eqnarray}
\nonumber V_1 \sum_{i} ~( n_{i} ~ n_{i+\hat{x}} + n_{i} ~ n_{i+\hat{y}}),
\end{eqnarray}
%%%%%%%%%%%%%%%%%%%%%%%%%%%%%%
is added to the asymmetric BHM. In the absence of hopping energies, NN interaction breaks the translational symmetry of the system and incompressible density-wave phase (DW) with the checkerboard pattern appears. Fillings factor of the DW phases increase with the chemical potential $\mu$ in every lobes. As $t/U$ increases, bosons can move coherently through the lattice  and supersolid phase emerges where both the translational and U(1) symmetries are broken simultaneously. At larger hopping energies, the translational symmetry is restored, and the system enters the superfluid phase. Asymmetric hopping energy frustrates bosonic motions and stabilizes solid order up to larger $t/U$ which enlarges the solid and SS phases.

From the above discussions, we expect that by frustrating hopping energy, bosons could not flow coherently which increases insulator borders. But note that involving magnetic field in the Eq. \ref{eq:model} is not just a simple changes of the hopping amplitudes in different directions, but the gauge field includes a phase factor in the hopping along one direction, such that hopping around a closed loop accumulates a nontrivial phase. 

%%%%%%%%%%%%%%%%%%%%%%%%%%%%%%%%%%%%%%%%%
%%%%%%%%%%%%%%%%% BHM+AMF %%%%%%%%%%%%%%%%
%%%%%%%%%%%%%%%%%%%%%%%%%%%%%%%%%%%%%%%%%

\section{Bose-Hubbard model with artificial magnetic field} \label{sec:BHM+AMF}

\begin{figure}[t]
\centering
\includegraphics[scale=0.42]{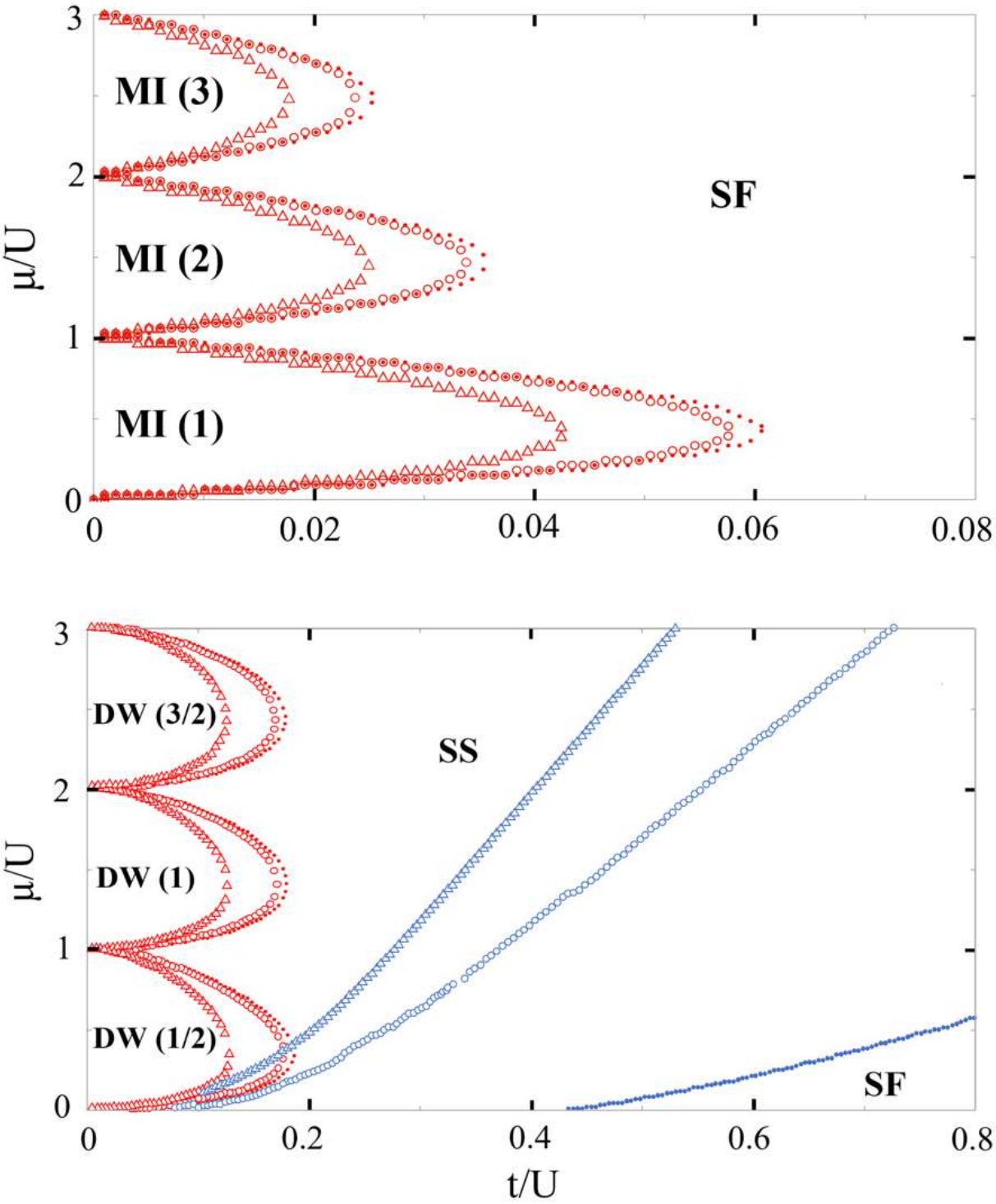}
\caption{SGMF $t-\mu$ ground state phase diagrams for different gauge fields $\alpha = 0$ (triangle), $\frac{1}{5}$ (circle) and $\frac{1}{2}$ (dot) in the units of on-site interaction ($U=1$) for top: BHM and bottom: eBHM with NN interaction $V_1/U=0.5$. MI-SF and DW-SS borders are plotted with red color and SS-SF borders with blue.}
\label{fig:PD-BHM+AMF}
\end{figure}

Using SGMF method, we solve BHM in the presence of artificial magnetic field for different ranges of gauge fields $\alpha$. Adding AMF to the BHM, explicitly breaks translational symmetry of the square lattice and the unit cell enlarges to $q \times 1$ for $\alpha = 1/q$. 
SGMF ground state phase diagrams of the BHM+AMF model are plotted in the top panel of Fig. \ref{fig:PD-BHM+AMF} for different ranges of magnetic field $\alpha$. 

%%%%%%%%%%%%%%%%%%%%%%%%%%%%%%%%%%%%%%%%
%%%%%%%%%%%%%%%%%%%%%%%%%%%%%%%%%%%%%%%
%%%%%%%%%%%%%%%%%%%%%%%%%%%%%%%%%%%%%%%
\begin{figure}
\centering
\includegraphics[scale=0.41]{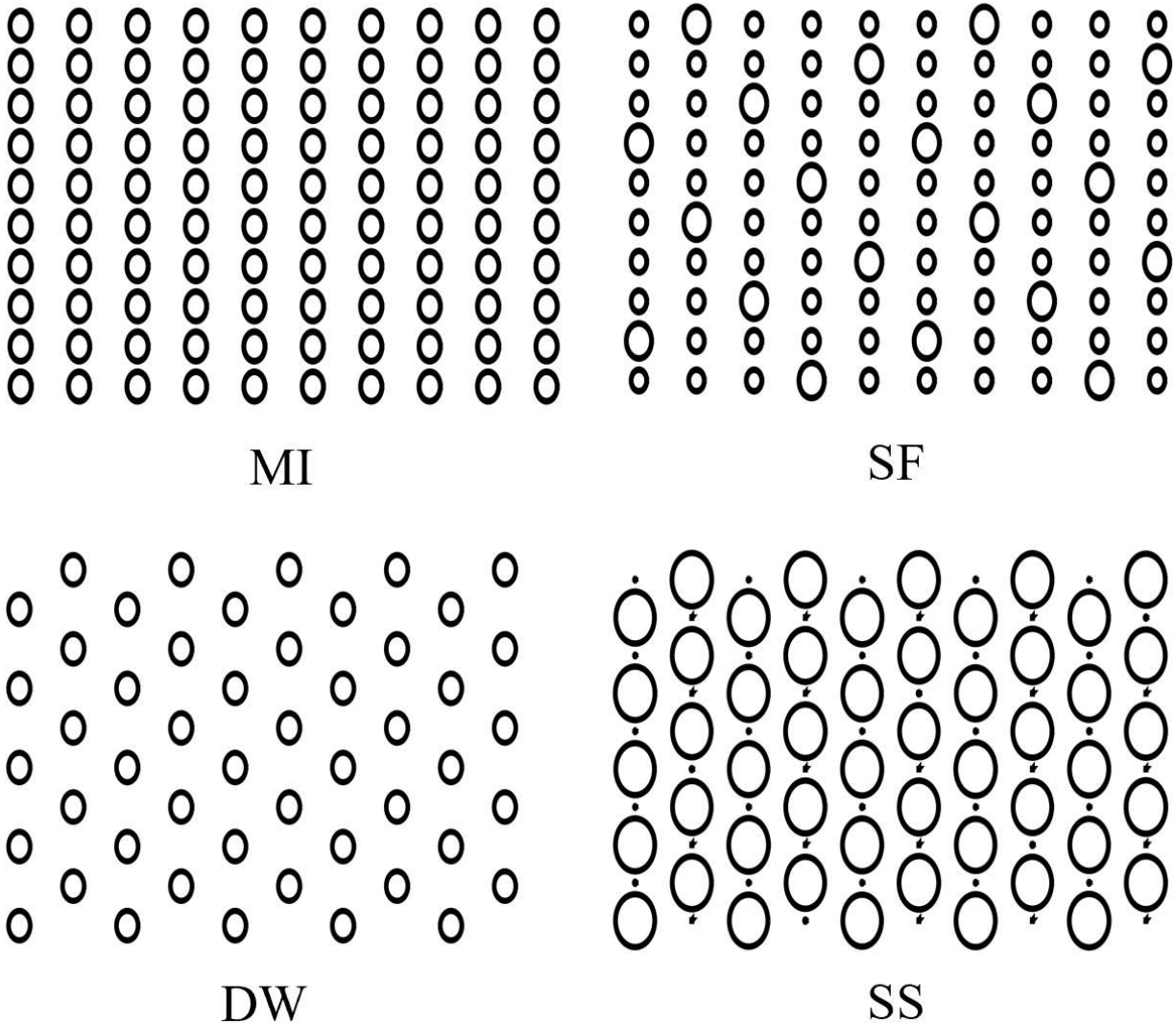}
\caption{Schematic SGMF plot of different phases for $\alpha = 1/5$ at various parameters ($t/U, \mu/U$): top left (0.01,1.5) in MI phase; top right (0.7,0.1) in SF phase; bottom left (0.1,1.5) in DW phase and bottom right (0.2,2) in SS phase. The radius of circles is proportional to the average number of bosons on every lattice sites and is unique in every frame. SF phase exhibits $q \times q$ pattern while SS phase has checkerboard one.}
\label{fig:phase-BHM+AMF}
\end{figure}
%%%%%%%%%%%%%%%%%%%%%%%%%%%%%%%%%%%%%%%%

In the absence of hopping energy, strong on-site interaction $U$ suppresses number fluctuations and eliminates the coherency between neighboring lattice sites. Therefore, incompressible MI phase appears where the original translational symmetry of the square lattice returns to the system.
In the Fig. \ref{fig:phase-BHM+AMF}, we have plotted schematic pictures of different phases. Filling factor of the MI phases increases with the chemical potential $\mu/U$ and their borders shrink with increasing hopping energy $t/U$. 

In the larger hopping energy, SF phase emerges where U(1) symmetry breaks simultaneously. Moreover, as the Fig. \ref{fig:phase-BHM+AMF} shows, the system undergoes broken translational symmetry in the both directions. The unit cell becomes $q \times q$ for $\alpha = 1/q$, something that was seen in the section \ref{sec:Ek} and is absent in the standard BHM. 

\begin{figure}[t]
\centering
\includegraphics[scale=0.415]{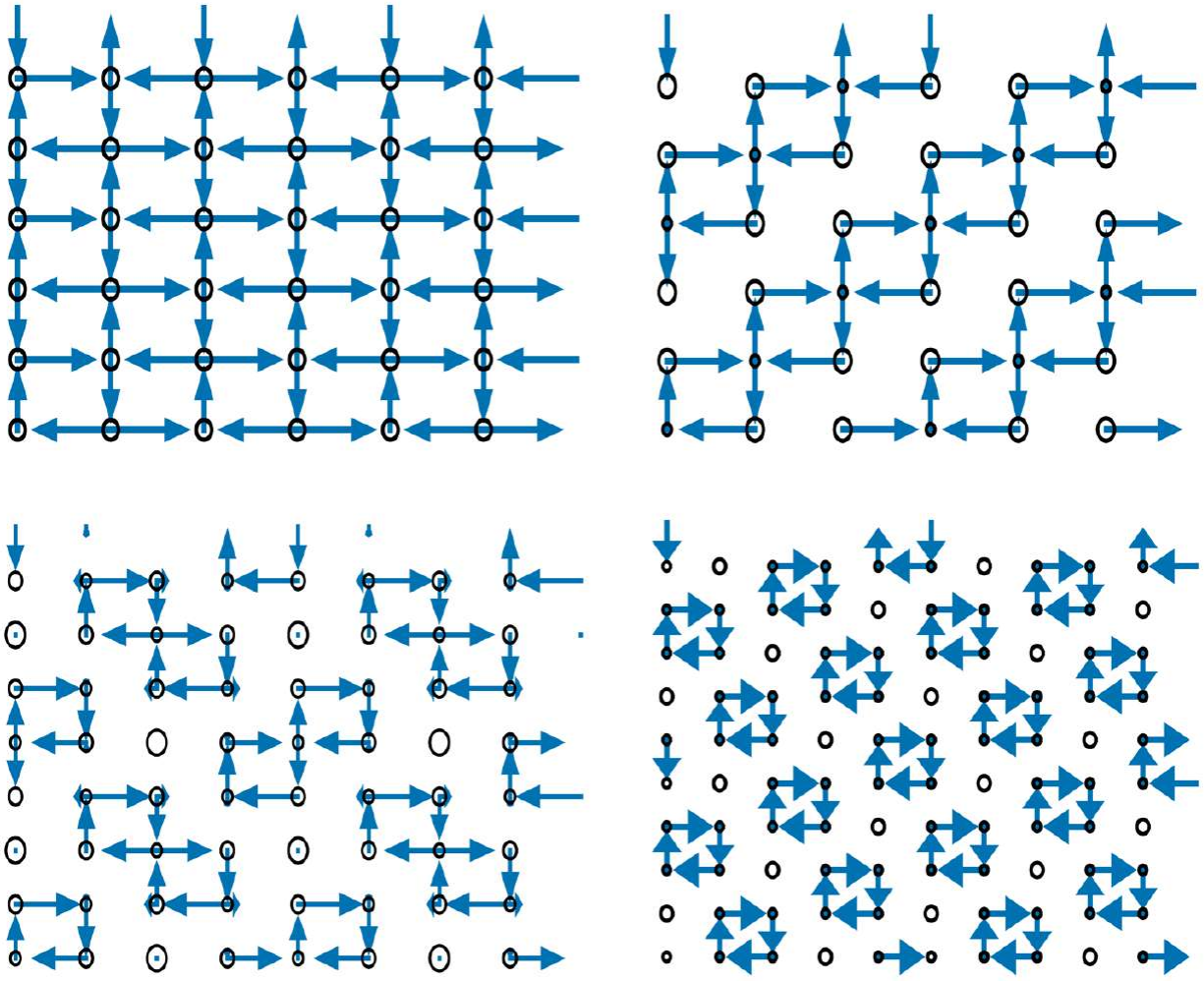}
\caption{SGMF bosonic currents (Eq. \ref{eq:j-op}) for the SF phase at $t/U = 0.8$, $\mu/U = 0.2$ and different gauge fields: $\alpha = \frac{1}{2}$ (top left), $\frac{1}{3}$ (top right), $\frac{1}{4}$ (bottom left), and $\frac{1}{5}$ (bottom right). The circles are positions of the lattice sites which their radius are proportional to the density $n_i$. The directions of arrows show the current $J_{i \to j}$ between lattice site $i$ and $j$. The length of the arrows indicate the magnitude of current in every subplots. The length scale is unique in every frame. Apart from  the $\alpha = \frac{1}{2}$, SS phase has similar patterns.}
\label{fig:current}
\end{figure}

Considering phase diagrams of the BHM+AMF with different ranges of $\alpha$, we find that insulator borders increase with the magnetic fields by localizing moving bosons via destroying coherency of bosonic motions. So as, the larger magnetic fields result the larger MI borders. Compare the $\alpha = 1/5$ and $1/2$ phase diagrams in the top panel of Fig. \ref{fig:PD-BHM+AMF}. 
 In order to find effects of gauge field on the bosonic motions of the condensed phases, we used the following directional operator known as bosonic current operator which calculate current from site $i$ to site $j$:
%%%%%%%%%%%%%%%%%%%%%%%%%%%%%%%%%%%%%%%
\begin{eqnarray} \label{eq:j-op}
J_{i \to j} = \mathrm{i} t ~ e^{\mathrm{i} \phi _{i,j}} ~ b^\dag_j b_i + h.c.,
\end{eqnarray} 
%%%%%%%%%%%%%%%%%%%%%%%%%%%%%%%%%%%%%%%
where $\phi _{i,j}$ measures the phase factor that bosons acquire by hopping from $i$ to $j$. According to the Eq. \ref{eq:model}, $\phi _{i,i+\hat{x}}=0$ and $\phi _{i,i+\hat{y}}=2 \pi  \alpha x_i$.
Bosonic current patterns for different ranges of AMF are plotted in the Fig. \ref{fig:current}. 

%%%%%%%%%%%%%%%%%%%%%%%%%%%%%%%%%%
%%%%%%%%%%%%%%%%%%%%%%%%%%%%%%%%%%%%%%%%

From the previous sections, including AMF leads to SF phases with simultaneous spatial orders in the both directions, which their configurations depend on the commensuration between the magnetic field and the lattice size. In the Fig. \ref{fig:current}, bosonic currents represent $q \times q$ periodicity and construct special vortex configurations for each $\alpha$. The number of vortices decreases and they become separate with $q$, which resemble Abrikosov lattices specially in the larger $q$ \cite{powell2011bogoliubov}. 

Using SGMF we consider effects of NN interactions in the BHM+AMF, see bottom panel of Fig. \ref{fig:PD-BHM+AMF}. In the absence of hopping energy, checkerboard solid orders appear for different ranges of $\alpha$. These DW phases persist up to larger $t/U$ where U(1) symmetry breaks spontaneously and supersolid order with broken translational and rotational symmetries appears. 
The filling factor and absolute value of the superfluidity exhibit checkerboard structure in the SS phase. Apart from  the $\alpha = \frac{1}{2}$, the current operator represents vortex patterns with $q \times q$ (for $\alpha = \frac{1}{q}$) pattern in this phase, see Fig. \ref{fig:current}. The bosonic current is vanished in the SS phase with the gauge field $\alpha = \frac{1}{2}$. 
Finally at large hopping energy, the SF phase appears. Filling factor and bosonic current has $q \times q$ (for $\alpha = \frac{1}{q}$) pattern in the SF phase, see Fig. \ref{fig:current}. 
As the bottom panel of Fig. \ref{fig:PD-BHM+AMF} shows, solid and SS borders increase with magnetic filed $\alpha$.

%%%%%%%%%%%%%%%%%%%%%%%%%%%%%%%%%%%%%%%%
%%%%%%%%%%% Finite temperature %%%%%%%%%%
%%%%%%%%%%%%%%%%%%%%%%%%%%%%%%%%%%%%%%%%%

\section{Finite temperature phase diagrams} \label{sec:temp}

%%%%%%%%%%%%%%%%%%%%%%%%%%%%%%%%%%%%%%%%%
\begin{figure}
\centering
\includegraphics[scale=0.41]{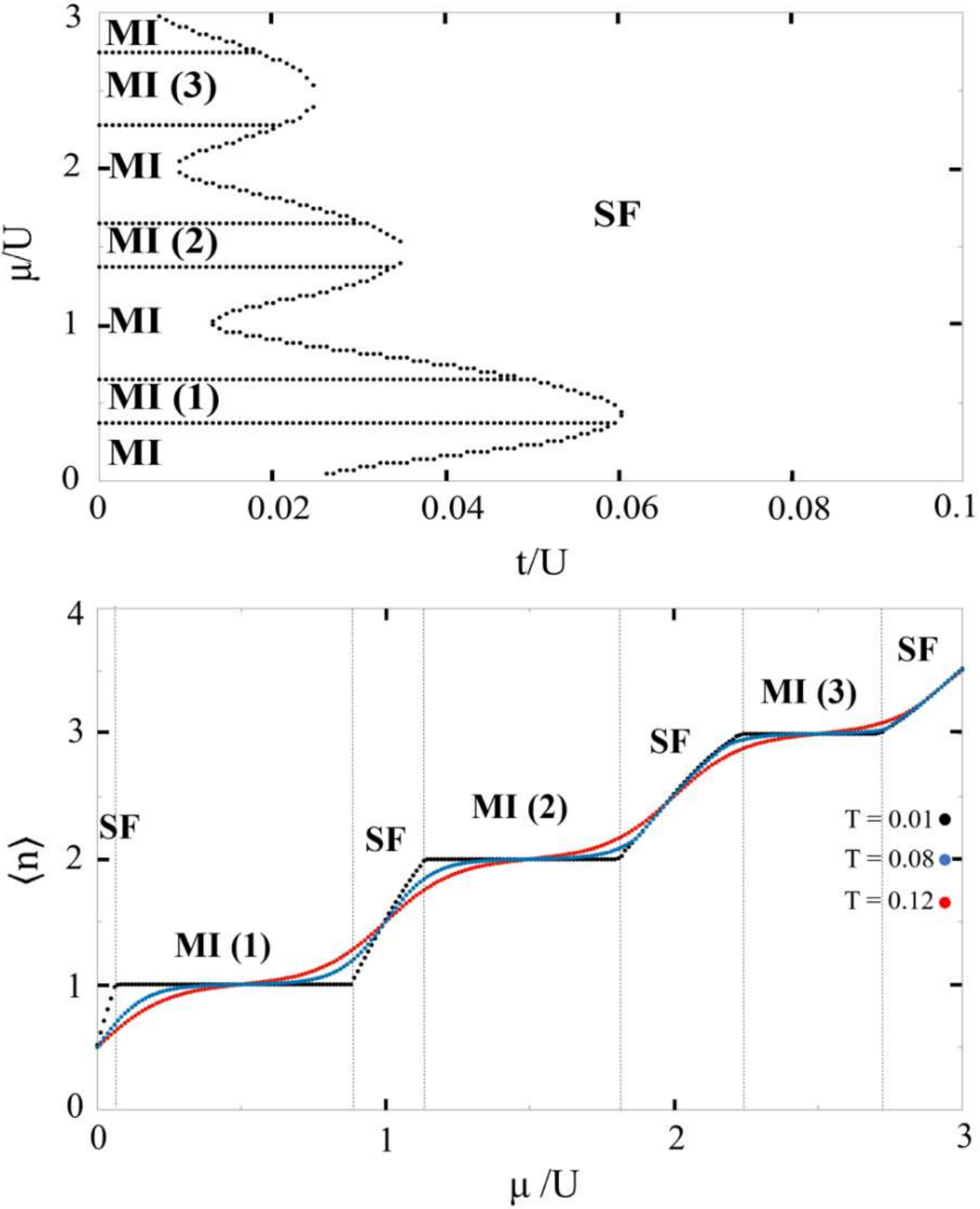}
\caption{Top: SGMF $t-\mu$ phase diagram for BHM+AMF at $\alpha = 1/2$ and temperature $T/U = 0.04$. Bottom: SGMF order parameter $\langle n \rangle$ for BHM+AMF at $\alpha = 1/2$, $t/U = 0.02$ and different temperatures.}
\label{fig:PD-BHM+AMF+T}
\end{figure}
%%%%%%%%%%%%%%%%%%%%%%%%%%%%%%%%%%%%%%%%%
%%%%%%%%%%%%%%%%%%%%%%%%%%%%%%%%%%%%%%%%%
\begin{figure} 
\centering
\includegraphics[scale=0.82]{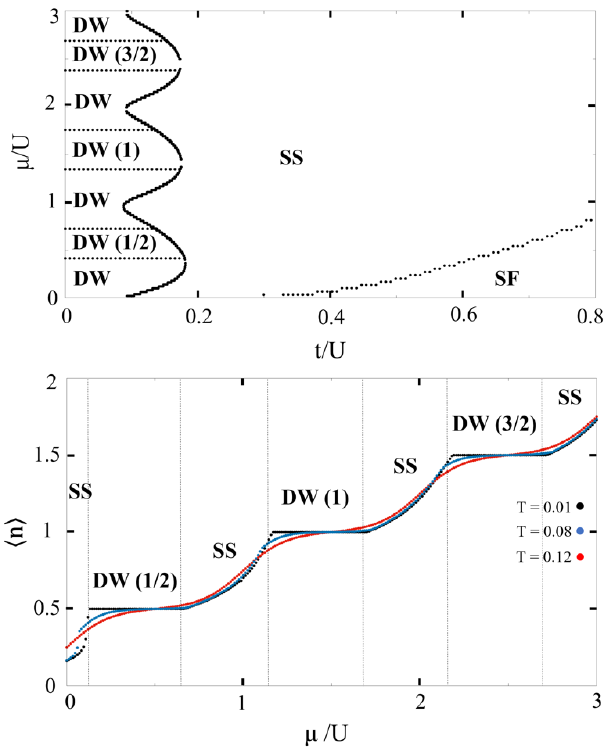}
\caption{Top: SGMF $t-\mu$ phase diagram for eBHM+AMF at $\alpha = 1/2$ and temperature $T/U = 0.04$. Bottom: SGMF order parameter $\langle n \rangle$ for eBHM+AMF at $\alpha = 1/2$, $V_1/U=0.5$, $t/U = 0.15$ and different temperatures.}
\label{fig:PD-eBHM+AMF+T}
\end{figure}

%%%%%%%%%%%%%%%%%%%%%%%%%%%%%%%%%%%%%%%%

At the ground state, there are long range phase coherence in the SF phase so as each boson is spread out over the entire lattice. On the other hand number fluctuation is zero in the insulator phases. 

In this section we consider thermal fluctuations on the ground state phase diagrams which is important for realizing BHM+AMF in experiments. At finite temperature $T$, thermal averages of the order parameters are used for determining phase diagrams. 
In the top panel of Fig. \ref{fig:PD-BHM+AMF+T}, we have plotted phase diagram of BHM+AMF at $\alpha=1/2$ and temperature $T/U=0.04$. 

As mentioned earlier, AMF localizes bosons and enhances the insulators and SS borders. Whereas thermal fluctuations delocalize bosons through the entire lattice and melt insulator phases. 
Because quantum fluctuations around the borders are larger, thermal fluctuations gradually decrease plateaus’ width from the borders and MI phase appears around the ground state insulator phases. In the MI phase both the translational and U(1) symmetries are present but average number of bosons is changed with $\mu$. 
Moreover, in the presence of thermal fluctuations superfluid density is suppressed and the system undergoes a transition to the thermal insulating phase with varying filling factor \cite{heydarinasab2017inhomogeneous, heydarinasab2018spin}, where both the translational and the U(1) symmetries are present. 

In the bottom panel of Fig. \ref{fig:PD-BHM+AMF+T}, average number of bosons is plotted for $\alpha=1/2$, $t/U=0.02$ and different temperatures $T$. 
Vertical lines determines the ground state borders and thermal fluctuations soften these transitions. 

\begin{figure}
\centering
\includegraphics[scale=0.41]{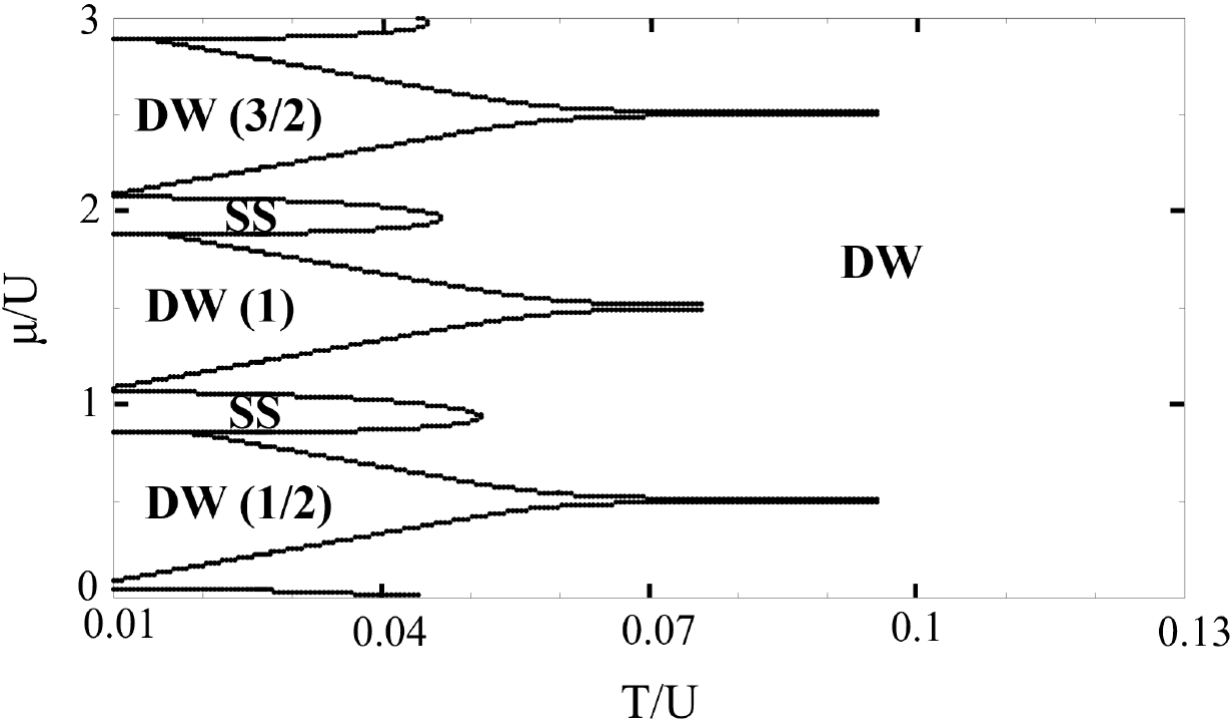}
\caption{SGMF $T-\mu$ phase diagram for eBHM+AMF at $\alpha = 1/2$, $V_1/U = 0.5$ and $t/U=0.1$.}
\label{fig:PD-T-mu}
\end{figure}

Thermal phase diagram of eBHM+AMF at $\alpha=1/2$, $V_1/U=0.5$ and $T/U=0.04$ is plotted in the top panel of Fig. \ref{fig:PD-eBHM+AMF+T}. 
DW phase appears around the plateaus as a result of thermal fluctuations, which has checkerboard structure but the average number of bosons is not constant. 
Bottom panel of this figure shows that melting starts from the borders (vertical lines). 
Thermal fluctuations decrease the SS phase by washing out the superfluid component but its borders are large enough to detect in experiments. At large $T$ which is comparable with interaction terms, the SS phase disappears and the system enters to DW phase, see Fig. \ref{fig:PD-T-mu}. 
Thermal fluctuations softens the transitions and melt the plateaus eventually at a transition temperature which depends on the strength of frustrations and interaction parameters. 
At large enough $T$ the system enters to the thermal insulator where all the symmetries return to the system and fillings increase with $\mu$.

%%%%%%%%%%%%%%%%%%%%%%%%%%%%%%%%%%%%%%%%%%%%

%%%%%%%%%%%%%% Conclusions %%%%%%%%%%%%%%
%%%%%%%%%%%%%%%%%%%%%%%%%%%%%%%%%%%%%%%%%

\section{Conclusions} \label{sec:conclusions}

We have considered the ground state and finite temperature phase diagrams of BHM and eBHM in the presence of AMF. 
At first, using asymmetric BHM and eBHM, we proved the enlargements of the insulators and supersolid phases at the result of the frustrated hopping energies. We related this increment to disruption the coherency of the bosonic motions. 
This was the case in the BHM and eBHM in the presence of AMF and the $\alpha=1/2$ had the largest regions. 
Moreover, the synthetic magnetic fields modulated the particles numbers and bosonic currents on the lattice sites. The spatial structure of these configurations depends on commensuration between the magnetic field and the lattice. 
In the MI phase translational symmetry of the square lattice returned while the DW and SS phases had checkerboard pattern. There are $q \times q$ configuration in the SF. Also the bosonic current exhibited vortex patterns with this symmetry in the SF and SS phases. 

Finally, we showed that the thermal fluctuations delocalize bosons and melts the plateaus and compressible phases appeared around the lobs. Also, superfluidity decreased with temperature. At large $T$ the compressible normal fluid appeared. 

It worth mentioning, although these results are obtained by mean-field theory, considering the previous studies \cite{suthar2020supersolid, pal2019enhancement} prove that quantum fluctuations is small enough to get ground state of Eq. \ref{eq:model} by SGMF qualitatively. 
This method gives quick initial results without limitations such as large computational cost of exact diagonalization methods. For example cluster Gutzwiller method needs large clusters ($q \times q$) to correctly include broken translational symmetry. 
Moreover, due to the infamous sign problem, quantum Monte Carlo cannot be performed in the presence of magnetic fields \cite{iskin2012artificial}.  
Nonetheless, including correlations between different sites with other methods results quantitatively more accurate borders. 

%%%%%%%%%%%%%% Acknowledgements %%%%%%%%%%%%
%%%%%%%%%%%%%%%%%%%%%%%%%%%%%%%%%%%%%%%%%%%%

\section{Acknowledgements}

The authors would like to thank Jahanfar Abouie and Abdollah Langari for useful discussions and reading the manuscript. Useful discussion with Saeed Abedinpour is acknowledged. 

\section*{Data availability
}
The data that support the findings of this study are available from the corresponding author (F.H.) upon request.

\bibliography{references.bib}

\end{document}